\documentclass{aa}
\usepackage{natbib,graphicx}

\begin{document}

\title{The visibility of the Galactic bulge in optical surveys. Application
  to the Gaia mission}

\author{A.C. Robin
        \inst{1}
        \and
        C. Reyl\'e \inst{1}
	\and
	S. Picaud \inst{1}
        \and
	M. Schultheis \inst{1,2}
   \institute{CNRS UMR6091, Observatoire de Besan{\c c}on, BP1615, 
    F-25010 Besan{\c c}on Cedex, France\\
\email{annie.robin@obs-besancon.fr,celine@obs-besancon.fr, picaud@obs-besancon.fr}
		\and
CNRS UMR7095, Institut d'Astrophysique de Paris, 98bis bld Arago, 75014 Paris,
        France\\
\email{mathias@obs-besancon.fr, schulthe@iap.fr}}
}
   \date{Received ; accepted }

   \titlerunning{The visibility of the Galactic bulge in optical surveys}

\abstract{The bulge is a region of the Galaxy which is of tremendous
interest for understanding Galaxy formation. However, measuring
photometry and kinematics in it raises several inherent issues, like
high extinction in the visible and severe crowding. 
Here we attempt to estimate the problem of the visibility of the bulge
at optical wavelengths, where large CCD mosaics allow to easily cover 
wide regions from the ground, and where future astrometric missions are 
planned. Assuming the Besan\c{c}on Galaxy model
and high resolution extinction maps, we estimate the stellar
density as a function of longitude,
latitude and apparent magnitude and we deduce the  possibility of 
reaching and measuring bulge stars. The method is applied to three Gaia 
instruments, the BBP and MBP photometers, and the RVS spectrograph.
We conclude that, while in the BBP most of the bulge will be
accessible, in the MBP there will be a small but significant number  
of regions where bulge
stars will be detected and accurately measured in crowded fields. 
Assuming that the RVS
spectra may be extracted in moderately crowded fields, the bulge 
will be accessible
in most regions apart from the strongly absorbed inner plane regions, because
of high extinction, and
in low extinction windows like the Baades's window where the crowding is
too severe.
   }

   \maketitle

\section{Introduction}

The
bulge is an important part of the Galaxy for understanding galaxy
formation and evolution, thus requiring surveys that go as deep as 
possible in this region. The bulge is highly extinguished
at its center and at low Galactic latitudes. It also suffers
from severe crowding when observed with low spatial resolution
instruments. The extinction is generally very
patchy. When extinction is high the crowding is less. Hence the
crowding in the
central regions of the Galaxy is very sensitive to the extinction
and may vary strongly from field to
field on a quite small spatial scale. 

Observations of bulge stars will strongly depend on the extinction and
on the crowding.
If the extinction is too high, the number of stars will be
low (no crowding) but conversely the bulge stars would not be observed, because
too faint. If the extinction
is low (like in the Baade's windows), bulge giants on the red clump 
are bright enough to be reached,
but the crowding will limit the number and/or the quality of their 
measurement. The number of stars in the bulge also strongly 
depends on latitude and longitude.
\cite{Drimmel2003b} attempted to localize the densest regions from the 
GSC-II catalogue. The densities, as estimated from this photographic 
catalogue, are given for all
populations together. They do not specify if bulge stars will be
accessible at the Gaia limiting magnitude G$\leq$20 or not.

Here we address the following question: is there a combination of 
parameters (extinction, latitude) in the Galactic bulge where the extinction is
large enough to avoid crowding and not too high to allow bulge star 
measurements in the optical ? In order to answer it, we have used the 
Besan\c{c}on Galaxy model \citep{Robin2003,Robin2004}
which allows us to simulate star counts as a function of magnitude, direction
and extinction. However these depend on the assumptions about 
Galactic structure, stellar populations and particularly the bulge model,
but even more strongly on the assumed extinction. 
 \cite{Schultheis1999}  have 
determined a detailed map of the mean extinction in the Galactic central 
region 
$|l|<9^\circ$ and $|b|< 1.5^\circ$ showing strong variations on a spatial 
scale of several arcminutes. The high patchiness is also visible in the
\cite{Schlegel1998} map
obtained from the dust column density deduced from FIR emission.

In Sect.~2 alternative extinction estimates are described. 
In Sect.~3 we present the
Galaxy model used for this study. We briefly describe 
the Gaia instruments in Sect.~4. 
In sect.~5 we show our estimation of the magnitude of bulge stars
reached as a function of the limiting G magnitude and of the extinction
map, and we address the problem of the crowding in Gaia instruments.
We also discuss the reliability of the results.

\section{Extinction in the Galactic central regions}

A fundamental parameter for estimating stellar densities in the bulge
is the extinction. If the extinction is too low,
the high stellar density will create heavy
crowding at the limiting magnitude. If the
extinction is too high
the bulge stars will be too faint to be observed.

Several extinction maps have been published in the past years 
\citep{Schlegel1998,Schultheis1999,Dutra2003,Drimmel2003a}. 
\cite{Schultheis1999} (hereafter SGS) have produced a detailed analysis of the 
extinction at a high
spatial resolution in the region -9$^\circ<l<9^\circ$ and -1.5$^\circ<b<1.5^\circ$.
This study was made using near-infrared photometry in J and K$_s$ bands 
from the DENIS survey \citep{Epchtein1997}. Isochrones of 10 Gyr and solar 
metallicity \citep{Bertelli1994} have been
fitted to J-K/K diagrams in 2 arcmin windows, by varying the extinction. 
The typical uncertainty is smaller than 2 magnitudes in windows with A$_V<$25. 
{ SGS assume a fixed
metallicity of Z=0.02. This corresponds to the mean value observed in red giants but a spread is known to be present in the
bulge. However the effect of not accounting for this spread should not
translate into a significant change in the total uncertainty.} 
The uncertainty due to the assumed age is also negligible. In regions
where A$_V>$25, the measurement is no longer reliable due to the lack of
detection in the J band. In this case the given value is an underestimate
of the true value. In small extinction regions (Av$\approx1$) the
accuracy is limited by the sensitivity of the extinction in the K band
and by the photometric accuracy which is about 0.05 in A$_K$, which
translates to 0.5 when computing A$_V$.
The errors in this extinction map 
increases towards higher extinguished regions due to the increase in
photometric errors. { The uncertainty on the extinction law may also
contribute to the total uncertainty, especially in high extinction regions}. 
\cite{Dutra2003} have shown  that their
map  based on 2MASS data agrees well with the DENIS extinction map 
\citep{Schultheis1999} up to $A_K \approx 1.5$ (about A$_V$=15).  
Above this limit,
the comparison is affected by increased internal errors in both
extinction determinations. However, a comparison with the Galactic
center
map of \cite{Catchpole1990} gives a good agreement with differences less
than A$_V$=2 mag.
The resulting map has a spatial resolution of 2 arcminutes and is the
average total extinction along that line of sight to a distance
of 8 kpc from the Sun.

\cite{Schlegel1998} (SFD) have produced a whole sky extinction map
using FIR dust 
emission from IRAS and COBE/DIRBE. They obtain a
high resolution map (at the resolution of IRAS) of the dust temperature and
of the dust column density and calibrate the extinction estimator with
a sample of elliptical galaxies. They also assume that the distribution of
dust grain sizes is the same everywhere in the Galaxy and that the large grains
which are responsible for the FIR emission are in equilibrium with 
the interstellar radiation field. The former hypothesis is
only correct in diffuse medium. They also fit each line of sight with
a single color temperature, which may not be true in the Galactic plane.
The practical error on E(B-V) in SFD map is estimated to be 0.011 mag.
However this value does not account for systematic errors in regions where their
assumptions do not apply - for example in regions with high gas to dust ratios,
in outflows around OB associations \citep{Burstein2003}. Schlegel et al. (1998) also noted that for latitudes
less than 5 degrees their extinction determination is less reliable.

Comparing SGS and SFD maps in the Galactic plane
leads to systematic differences. 
The SFD estimation is the total extinction along the line of sight through
the entire Galaxy. So, in the plane, the total extinction would be
about twice the value undergone by bulge stars, the difference being due
to extinction on the far side of the bulge. In general in the bulge we expect
SFS
values to be a significative overestimate of the extinction. However at higher
Galactic latitudes, this overestimate should become smaller.
Moreover, because the lines of sight in the Galactic plane may encounter
regions with different density and color temperatures, the SFD map is
probably less
reliable for estimating the extinction that bulge stars undergo. We have 
chosen to rely on the SGS
map in regions where it exists ($|b|<1.5^\circ$), and on the SFD map at
higher latitudes.

\section{The Besan\c{c}on Galaxy model}

In order to estimate the total number of stars, and the
number of bulge stars, in given directions as a function of magnitude,
we have used model predictions from the Besan\c{c}on Galaxy model 
\citep{Robin2003, Robin2004}. We describe below the main ingredients 
relevant to the present study.

\subsection{Overview of the stellar population model}

The model is based on a population synthesis scheme. Four distinct
populations are assumed (a thin disc, a thick disc, a bulge and a spheroid)
each deserving specific treatment. In the central regions, only
the bulge and thin disc are important, the thick disc and spheroid being
negligible at the magnitudes considered here.

The thin disc is a prominent population in the bulge region, especially 
at optical
wavelengths. {Although the bulge density exceeds the disc density at 
Galactocentric distances smaller than 3.14 kpc, the bulge is less visible
due to extinction. Hence it is necessary to adequately model the
foreground disc population. 
The density is modeled by a Einasto law with a scale length of 2.35 kpc and
it develops a progressive central hole of scale length 1.32 kpc
\citep{PicaudRobin2004}. The maximum density of the disc is at about 2 kpc from
the center.
The disc population} is described by an evolutionary scheme with a constant star formation rate
over the past 10 Gyr. In addition, a two slope IMF has been used with a high 
mass slope (alpha=3), slightly
higher than Salpeter's one \citep{Haywood1997}.

\subsection{The outer bulge}

\cite{Picaud2003} has undertaken a detailed analysis of the outer bulge stellar
density and luminosity function by fitting model parameters to a set
of 94 windows in the outer bulge situated at -8$^\circ<l<10^\circ$ and 
-4$^\circ<b<4^\circ$. The data were obtained by the
DENIS survey team (Simon et al. in preparation) using K$_s$ magnitude
and J-K$_s$ color distributions. Using a
Monte Carlo method to explore a 11 dimensional space of bulge and disc 
density model parameters
and a maximum likelihood test of goodness of fit, he showed that the
bulge follows a density law which can be modeled by a boxy exponential or 
a boxy sech$^2$ profile. He found a triaxial bulge with the major axis
pointing towards the first quadrant with an angle of about 10$^\circ$ 
with respect to
the Sun-Galactic center direction. A full description of the  parameter 
values of
the bulge density law can be found in \cite{PicaudRobin2004}. Several 
luminosity functions have also been tested. The favoured combinations
between age and evolution model are
the \cite{Bruzual1997} one with an age of 10 Gyr, or the 
\cite{Girardi2002} one with an age of 7.9 Gyr.

\begin{figure*}
\centering
\includegraphics[width=12cm,clip=,angle=-90]{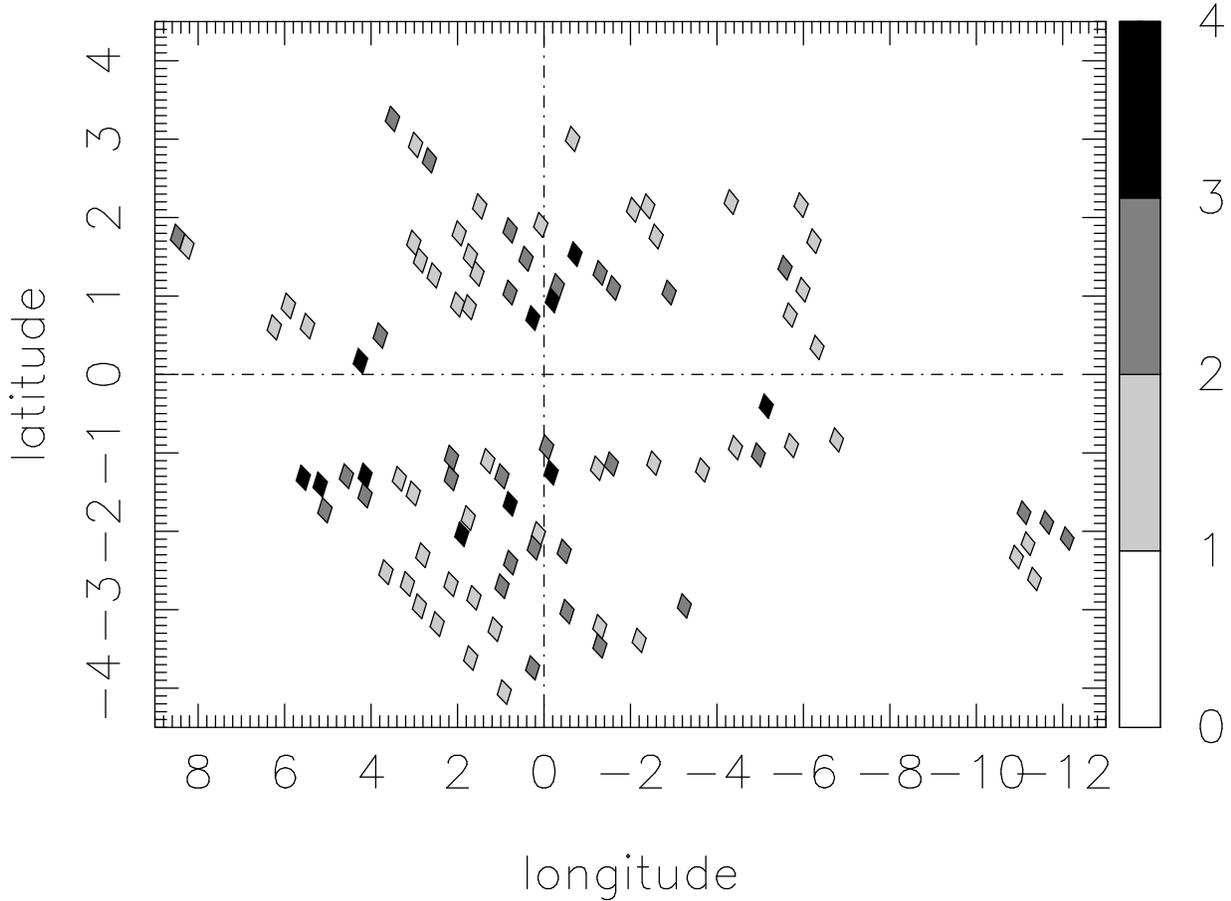}
   \caption{Overall goodness of fit of the model in 94 windows with 
NIR data. The greyscale is a function of the square root of the
$\chi^2$ per bin of magnitude-colour in each window. Two regions are
found at more than 3 sigmas: the first one near the bulge minor axis 
at $l\approx 0^\circ$, $b\approx 1^\circ$, the second one
at $l\approx 5^\circ$, $b\approx -1.5^\circ$.}
   \label{fig1}
\end{figure*}

In figure~\ref{fig1} we show the overall goodness of fit of the resulting model
with a \cite{Bruzual1997} 10 Gyr luminosity function. The figure looks similar
using \cite{Girardi2002} tracks, as the worse windows are the same for both
stellar models. The greyscale code is a function of
the square root of the $\chi^2$ per bin of colour-magnitude.
The overall agreement is good in most of the windows except in two regions:
the first one close to the Galactic center ($l=0^\circ, b=+1^\circ$), 
an overdense region with regard to the Galaxy model which is
unaccessible to optical surveys because of very high extinction; the second 
one being situated at around $l\approx5^\circ$ and 
$b\approx-1.5^\circ$. The
latter discrepancy is probably due to 
a bad extinction distribution model specifically in these windows.

We will use this model to
estimate the number and the absolute magnitude of bulge stars
as a function of position, apparent magnitude and interstellar extinction.
We also compute for the same parameters the total number of stars 
to address the problem of crowding.

To apply the method to Gaia instruments we have computed the
G magnitude from the formula defined for the new design
 of the payload (Gaia-2):

\[G= V-0.00075 - 0.3757\times (V-I) - 0.1233\times (V-I)^2 + 0.0061\times (V-I)^3 \]

The formula is applied to V and I magnitudes reddened following the \cite{Mathis1990} extinction law.

\begin{figure*}
\centering
\includegraphics[width=4.2cm,angle=-90,clip=]{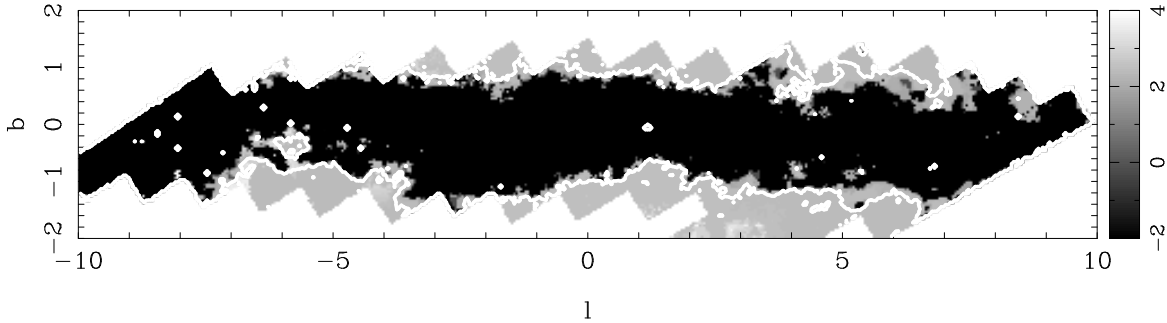}
   \caption{Absolute magnitude M$_V$ of bulge stars just reached at 
magnitude G=17, as a function
of latitude and longitude, according to the \protect\cite{Schultheis1999} 
extinction map. An absolute magnitude of -2 means that no bulge stars
are reached. The solid contour shows the iso-density of 20,000
stars per square degree, which is the crowding limit of the RVS. The
regions which do not suffer from crowding with the RVS are
between the dark dust lane and the iso-density contour.}
   \label{figschul17}
\end{figure*}

\begin{figure*}
\centering
\includegraphics[width=12cm,angle=-90,clip=]{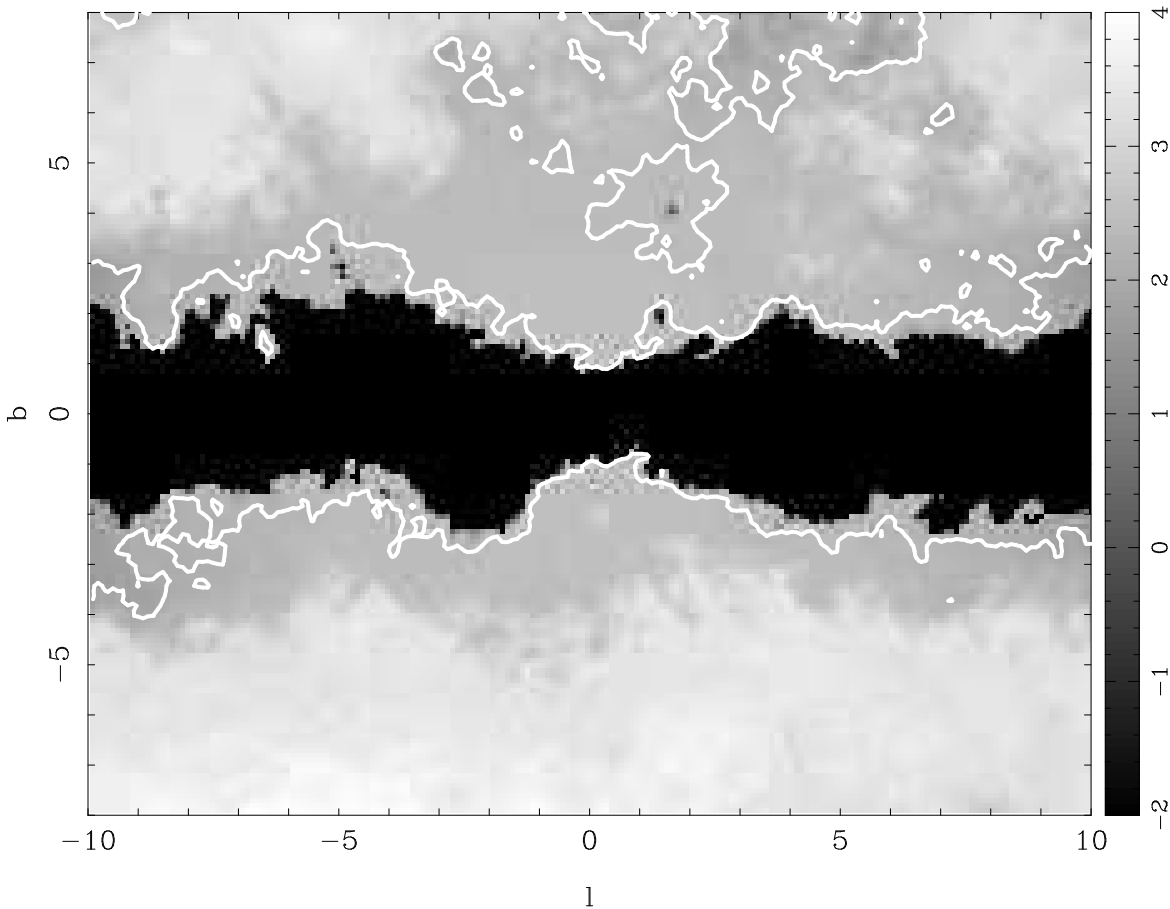}
   \caption{Same as figure~\ref{figschul17} but using the
\protect\cite{Schlegel1998} extinction map. {The white lines are isodensity
contours at 20,000 total stars per square degree, the assumed crowding
limit of the RVS}. }
   \label{figschleg17}
\end{figure*}
\begin{figure*}
\centering
\includegraphics[width=12cm,angle=-90,clip=]{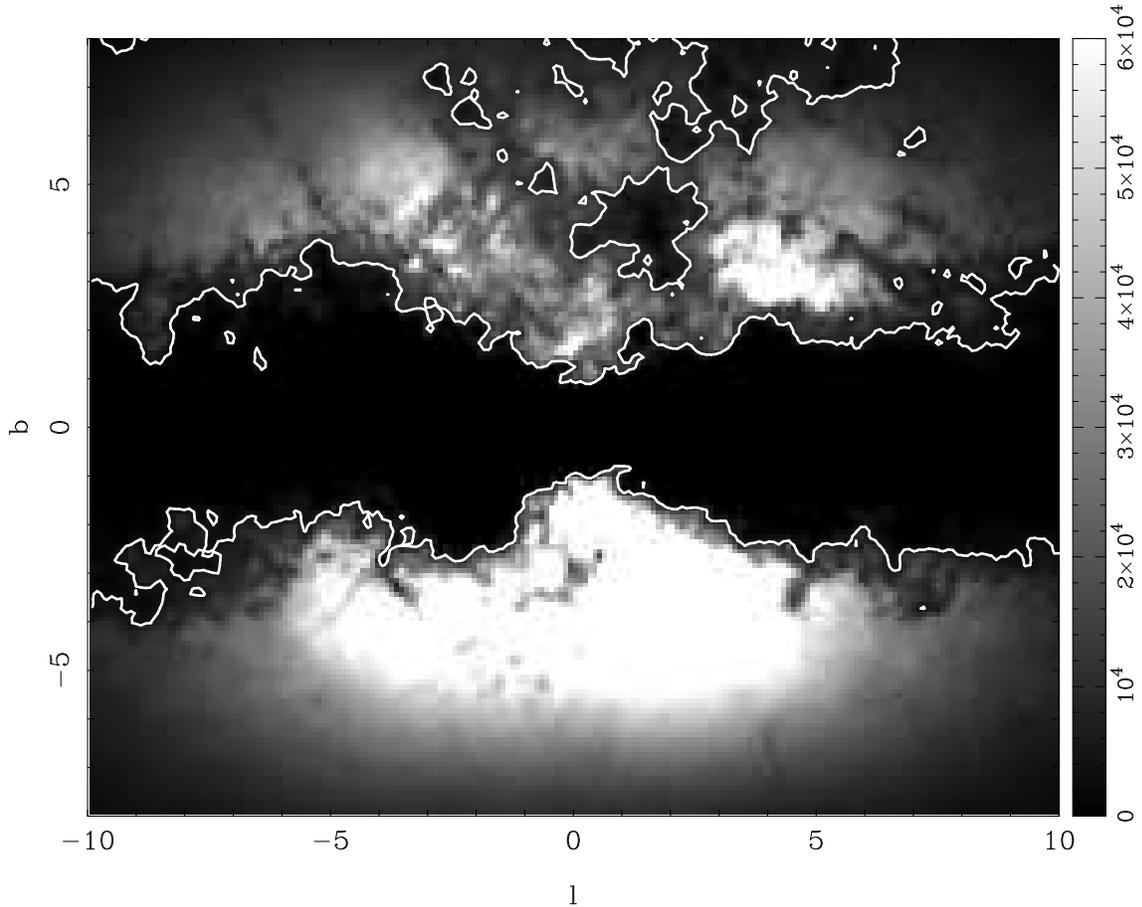}
   \caption{Bulge stellar density in stars per square degree at G $<17$ 
assuming the
\protect\cite{Schlegel1998} extinction map. The white lines are isodensity
contours at 20,000 total stars per square degree, the assumed crowding
limit of the RVS. }
   \label{figschleg17bul}
\end{figure*}

\begin{figure*}
\centering
\includegraphics[width=12cm,angle=-90,clip=]{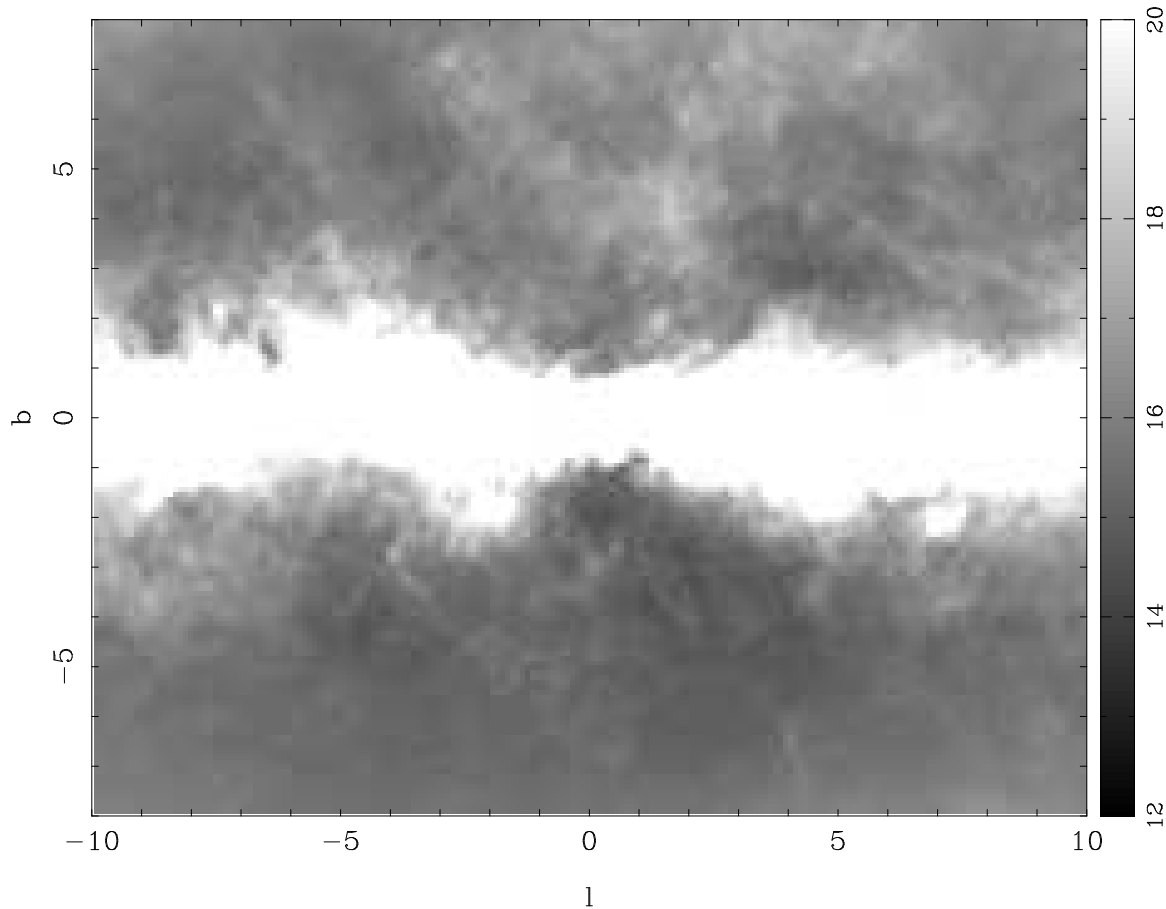}
   \caption{Apparent G magnitude at which the RVS crowding density of
 20,000 stars per square
degree (bulge+disc) is reached, assuming the \protect\cite{Schlegel1998} 
extinction map.}
   \label{schlegmag2}
\end{figure*}

\begin{figure*}
\centering
\includegraphics[width=12cm,angle=-90,clip=]{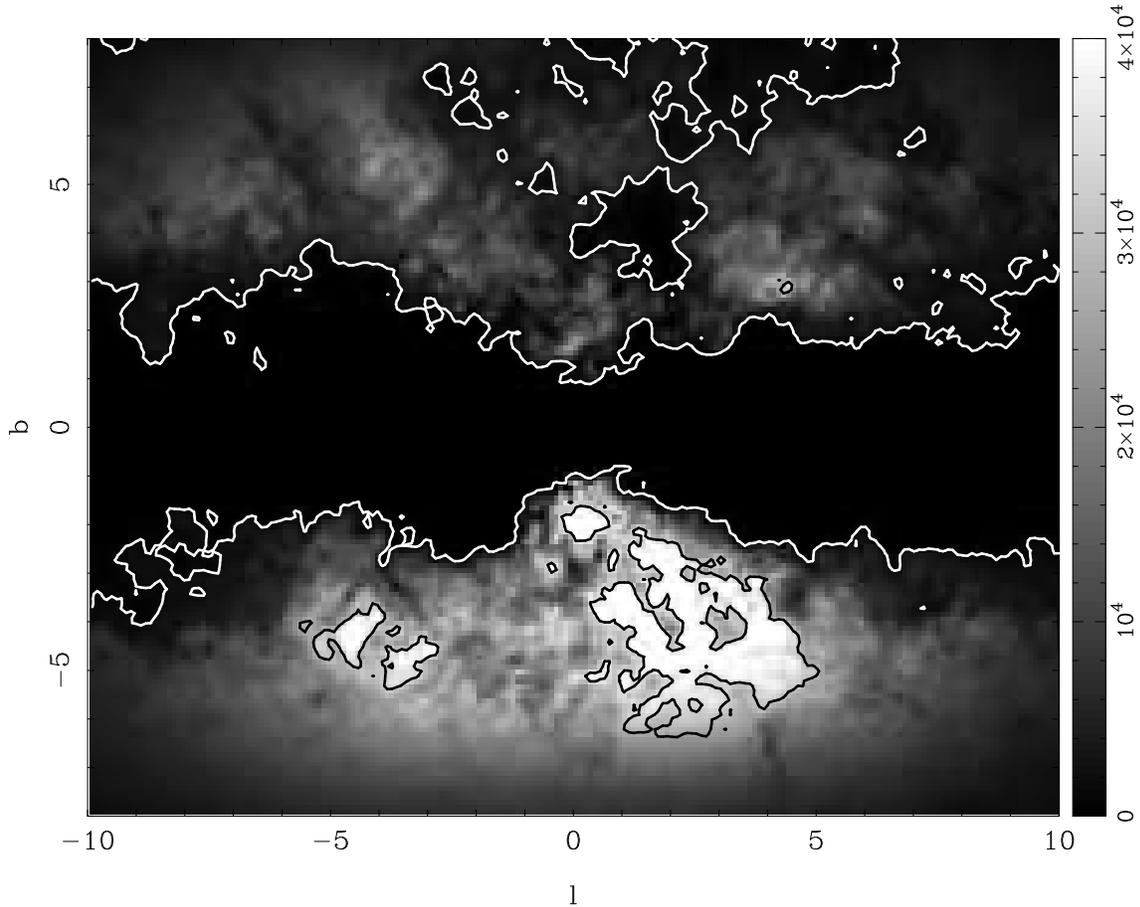}
   \caption{Bulge stellar density in stars per square degree
at magnitude
G $<15$, assuming the \protect\cite{Schlegel1998} extinction map. The white lines 
are the isodensity contours at 20,000 stars per square
degree (bulge+disc) at G$<17$, while the black lines are the same
at G$<15$. Less extinguished regions at b$<$-1 (inside black contours) suffer
from crowding even at G$<$15.}
   \label{figschleg15}
\end{figure*}

\subsection{Reaching the bulge at magnitude 20 with the photometers}

\begin{figure*}
\centering
\includegraphics[width=4.2cm,angle=-90,clip=]{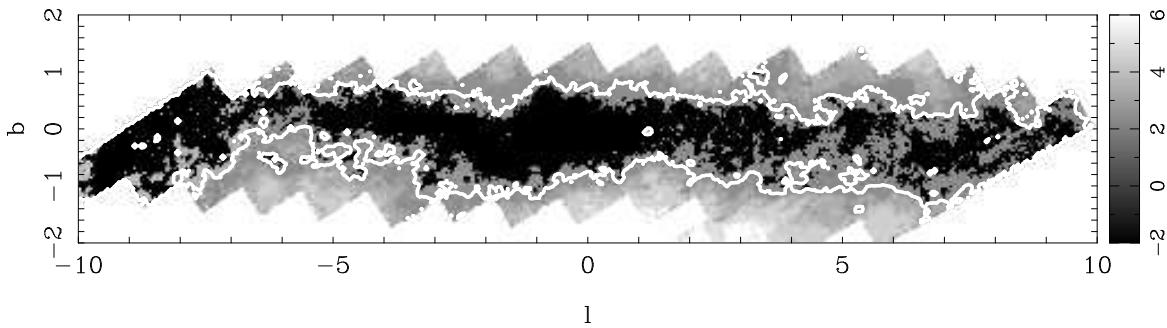}
   \caption{Absolute magnitude M$_V$ of bulge stars just reached at 
magnitude G=20, as a function
of latitude and longitude, according to the \protect\cite{Schultheis1999} 
extinction map. An absolute magnitude of -2 means that no bulge stars
are reached.
The solid contour shows the iso-density of 100,000
stars per square degree, which is the crowding limit of the MBP. The
region which is accessible to observation with the MBP is placed in
between the dark dust lane and the iso-density contour.}
   \label{figschul20}
\end{figure*}

\begin{figure*}
\centering
\includegraphics[width=12cm,angle=-90,clip=]{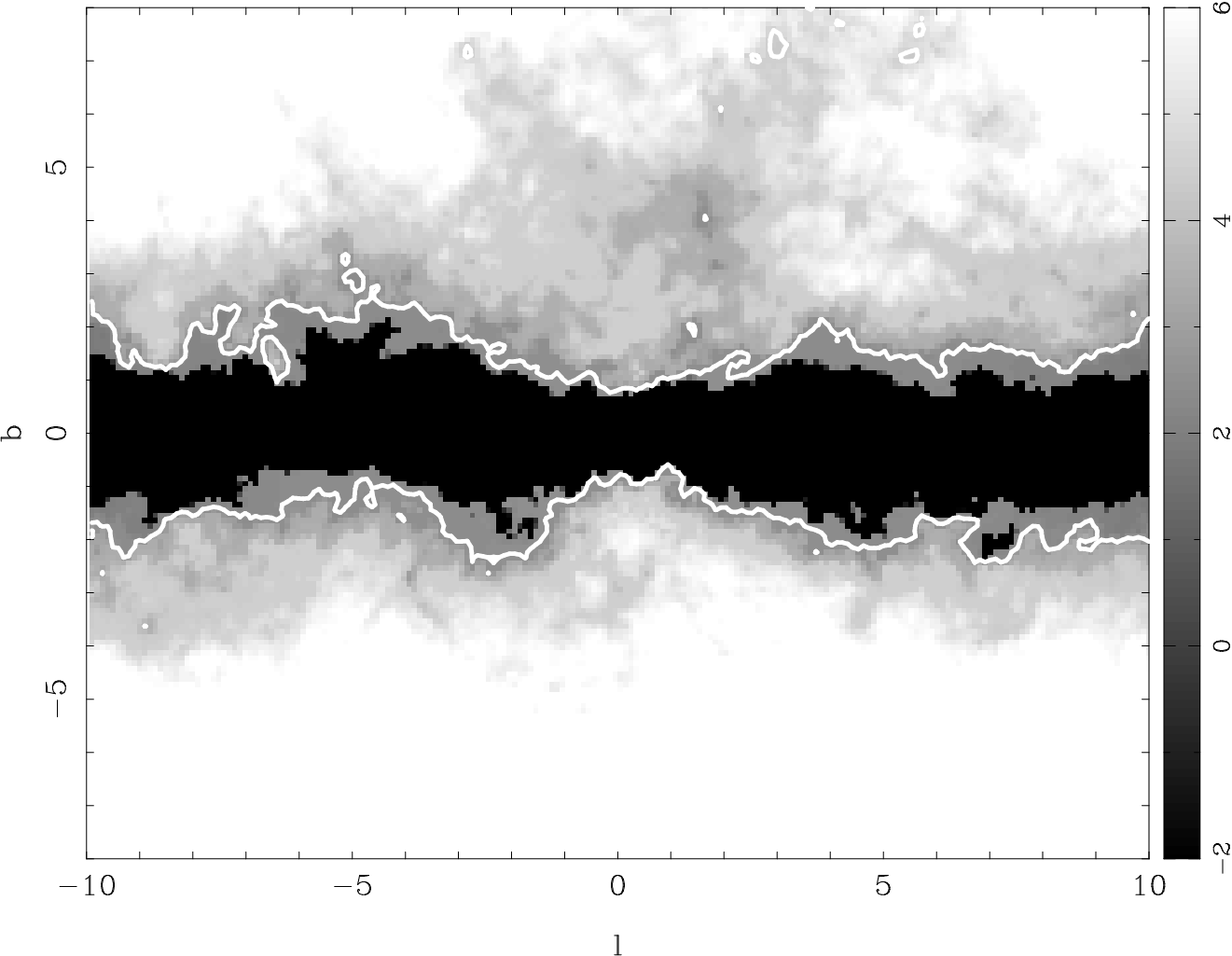}
   \caption{Same as figure~\ref{figschul20} but using 
\protect\cite{Schlegel1998} extinction map.}
   \label{figschleg20}
\end{figure*}

Figure~\ref{figschul20} and \ref{figschleg20} are similar to
Figures~\ref{figschul17} and \ref{figschleg17} but at the limiting
magnitude G=20 and for a different crowding limit of 100,000 stars per
square degree. 

\section{Crowding in the Gaia instruments}

The crowding in Gaia instruments depends on the sensitivity and on the
spatial resolution of each instrument. We shall consider each one separately.

\subsection{Broad Band Photometer (BBP)}

The astrometric instruments will provide accurate photometry in a few wide
bands with the Broad Band Photometer (BBP) and astrometry in the
Astrometric Field (AF).
The pixel and sample angular areas of BBP are
44.2$\times$132.6 and 44.2$\times$1591.2 mas$^2$, respectively.
The instrument qualifications have been chosen in order that it rarely
reaches crowding, to be able to measure up to about 3 million 
stars per square degree \citep{Jordi2002}. This limit is imposed by the 
number of windows around each
star that can be downloaded according to the telemetry
budget and to the size of these windows. This may be adapted but at
the price of a loss of astrometric and photometric precision at the
limiting magnitude.

The sensitivity of the instrument will allow completeness
upto the magnitude of about G=20. 
There may be a few regions in the Galactic plane and in 
the center of some of the
globular clusters reaching this density limit (3.10$^6$ stars
per square degree).

\subsection{Medium Band Photometer (MBP)}

The Medium Band Photometer (MBP) instrument is dedicated to obtaining 
accurate medium band
photometry in about 11 filters up to magnitude 20. The angular size of
the pixel and the sample are
1$\times$1.5 arcsec$^2$ and 1$\times$6 arcsec$^2$, respectively. This means 
that at a density of
about 270 000 stars per square degree the CCD will be completely full of stars.
The crowding limit has been estimated to be between 50,000 and 100,000 stars 
per square degree
\citep{Hoeg2002} but this estimate is still under more detailed 
investigation. {When the crowding is large in the MBP it is difficult
to cross-identify a star detected in the AF and BBP with a measurement in the MBP, even for bright stars, several stars being superimposed}. In the present study we have assumed a 
crowding limit of 100,000 stars per square degree.

\subsection{Radial Velocity Spectrograph (RVS)}

The Radial Velocity Spectrograph (RVS) instrument aims at observing a wavelength interval of about
26 nm around $\lambda_c$ = 861.5 nm. The spectral resolution has been recently 
chosen to be R=11500. The spectra will be 694 pixel long and
3 pixel wide (100\% energy).
During the Gaia mission, each target will be observed 102 times on
average. Due to
the scanning law, each time the orientation will be different so that if ever
the star overlaps its spectrum with a neighbouring star
it will not be the case in the next transit, at least not with the same
star. However, in very dense fields the overlaps can happen in a significant
number of transits for a given star.

\cite{ZwitterHenden2003} have estimated the probability of overlapping
between two neighbouring stars
as a funtion of stellar density. The overlap appears when there is 
more than one star in 694 pixel length. For a stellar density of 1,200 
stars per square degree at V$<$17 
they found that about 21 transits in 100 are subject to 
overlap. Going to a density of 6,000 stars per square degree increases
the fraction of overlap to about 70\%.

In a complementary study
Zwitter (2003) estimated how the accuracy on radial velocities may 
depend on the crowding. 
He argues that the overlaps degrade the radial velocity
accuracy only at faint magnitudes (V$\>=$17.5) and in highly crowded fields.
This is mainly because estimates of the overlapping spectra can be
made during the Gaia survey: photometric broad band photometry and accurate
astrometry will allow a first order correction of the overlapping spectra
for each transit, 
knowing the accurate position and magnitude,
and an estimate of the spectral type of the overlapping star.
However the recovering of other astrophysical parameters (temperature, 
gravity, metallicity) will be limited to bright targets in not too dense
environments.

In estimating the possibility of observing bulge stars with the RVS, 
we have assumed a crowding limit of 20,000 stars per square degree,
following
recommendations from the RVS Working Group (Katz, private
communication) and a
limiting magnitude of G=17.

\section{Reaching bulge stars}

Using the Besan\c{c}on Galaxy model and alternatively the 
SGS and SFD extinction maps, we are able to
estimate the magnitude 
of bulge stars reachable as a function of the limiting magnitude.
Thanks to the 3D modeling, bulge stars are considered in depth, their
distances ranging between 6 and 11 kpc on the line of sight.
Therefore in some cases only the close side of the bulge is visible
while in more favourable cases stars throughout the bulge will be observable.

As an example we have applied this method to the
Gaia instruments, the limiting magnitude being G=17 for the RVS 
and G=20 for the BBP and MBP 
photometers. The mean absolute M$_V$ magnitude of bulge clump giants is 
about 0.75 but
their luminosity function rises rapidly on the red giant branch
starting at about M$_V$=0. In the following, we consider that the
bulge is reached
when a significant number of giants are above the limiting magnitude.
We have set this limit at 100 bulge stars per 
square degree.

The density strongly depends on the assumed extinction.  We have 
used the two maps described above, knowing that the SGS map
is probably free from systematic errors and more reliable at low Galactic
latitudes, and the SFD map is suitable at higher latitudes. In the 
following we preferentially use the SGS map at latitudes $|b|<1.5^\circ$
and the SFD map and higher latitudes (in absolute values). {In both cases, the
absorbing clouds are assumed to be located at about 1 kpc from the Sun
in the first spiral arm.}

\subsection{Reaching the bulge at magnitude 17 with the radial velocity spectrograph}

In the RVS instrument of Gaia, the limiting magnitude is estimated at G=17.
Figures~\ref{figschul17} and \ref{figschleg17} give the absolute magnitude 
of bulge stars reached at the limit of 100 bulge stars per 
square degree at G=17. Figure~\ref{figschul17}
assumes the extinction from the SGS map and figure~\ref{figschleg17} from the SFD map.

In these figures, an absolute magnitude equal to -2 means that no
bulge stars are reached. 
The regions concerned are the
very low latitudes $|b|<1^\circ$. Going to latitudes around 1-1.5$^\circ$,
according to the SGS map there are regions where a number of bulge stars
of the clump will be accessible (regions where magnitudes M$_V$ between 0 
and 2 are reached). If one relies on the SFD map, those regions may lie
at about $|b|=1.5$ to $3^\circ$ depending on longitudes.
These regions will be very interesting to observe with the RVS. However
they may suffer from heavy crowding. In order to check this point, we 
have superimposed on 
figures~\ref{figschul17} and \ref{figschleg17} the limiting contour of stellar
density of 20,000 stars per square degree.
The disc dust lane prevents observations at very low latitudes. On the other
hand the crowding will cause observing problems at higher latitudes when
the absorption is small enough. Between these, there are at each longitude, on each side of the dust lane, fields where bulge stars
will be observable in not-crowded fields with the RVS. In the figures 
these fields lie 
between the dark pixels (where no bulge stars are detected) and the 
contour (at the limit of crowding of 20,000 stars per square degree). Depending
on the assumed extinction map they are at
latitudes of about 1-1.5$^\circ$ (SGS map) or a bit above at 1-3$^\circ$ (SFD
map). The first number should be prefered as, at these low latitudes, SFD maps
are less reliable. 

There are also a number of windows at slightly higher latitudes (for
example at b=4$^\circ$, l=1$^\circ$ and at  b=6-8$^\circ$,
l=1-6$^\circ$) which avoid the crowding in the RVS.

{Figure~\ref{figschleg17bul} shows the number density (per square degree)
of bulge stars at G$<$17, hence are measurable with the RVS. The numbers 
are very high, apart
from the regions very close to the plane, where the extinction is too heavy
to allow the bulge clump to be reached at this G magnitude.}

{In crowded regions, the RVS may still obtain useful spectra but limited 
to brighter magnitude stars. Figure~\ref{schlegmag2} shows the apparent 
G magnitude
at which the total density of 20,000 per square degree is reached 
(bulge and disc) assuming the SFD extinction map. On either side of the 
low latitude region heavily obscured by dust, the magnitude of the crowding
ranges from 14 to 18. 
Most regions will not suffer from crowding at magnitude G=15, apart from
a region at negative latitudes, at longitude around zero, and patchy regions
at positive latitudes.

Assuming that in most regions at G $\leq$ 15, the density is smaller than
the crowding limit, we can estimate how many bulge stars may be measured 
at G $\leq$ 15 
with the RVS and 
where. We show in Figure~\ref{figschleg15}
the number of bulge stars observable with the RVS at G$<15$. The 
white lines indicate the limits of the total density of stars of 20,000
per square degree
at G$=17$ and the black contours the same density limit but at G=15. 
The regions encircled by the black lines are in heavily crowded regions (among them is Baade's window) where the
RVS will obtain reliable spectra only for the brightest stars, which could 
be contaminated by
the high background. In the regions
between the black contours and the white ones, the crowding will be limited
at the magnitude interval 15$<G<17$.}

The position of the
fields given here should be taken with caution because of the extreme
sensitivity of the computation to the extinction, which is not known 
 with an accuracy better 
than about 2 magnitudes in these obscured regions.
However, if the extinction map does not suffer from systematic errors, 
there should be regions with appropriate
numbers of stars and extinction, allowing the observations of the bulge
stars at any longitude quite close to the dust lane.

The absolute magnitudes of detected bulge stars will range between 1 and 2 in
M$_V$ allowing study of the kinematics and metallicities of  bulge giants as
a function of longitude, a very good test for the bulge formation scenario.

At low latitudes (figure~\ref{figschul20}) the dust lane masks many fields
where even the brightest bulge stars will not be reached at G$<$20.
However there remain a number of fields even at latitudes close to 0$^\circ$ 
where one can reach bulge stars in the giant clump (M$_V~\approx~1$).

At higher latitudes ($|b|>1-2^\circ$ from figure~\ref{figschleg20}) all fields 
will be accessible. However depending on the instrument, these stars may be
in crowded fields. In the BBP photometer and AF astrometer there will
be no crowding
problem, the density of stars never reaching the value of 3 million stars
per square degree.
In the MBP , however, the crowding will prevent one from observing bulge stars
in a large number of cases, as shown by the iso-density contour at
100,000 stars per square degree superimposed in the figures.

Only in a small range of intermediate latitudes (fig~\ref{figschul20}, see the contour) will the bulge stars
be accessible on each side of the dark dust lane at $|b|\leq 1^\circ$.
These stars have absolute magnitudes of about 1-2 and are mainly clump
giants. At 1-2$^\circ$ the extinction goes down and the number of 
stars increases very rapidly to over 100,000 stars per square
degree. If the SFD map was used in place of the SGS map (fig~\ref{figschleg20})
the range of latitudes where bulge stars will be accessible to the MBP
would be slightly different and a bit more optimistic. It roughly ranges 
between $1<|b|<2^\circ$. However, as discussed above, the SFD map may be unreliable
at these latitudes.

Higher latitude
regions will barely be measured with the MBP due to crowding. 
{Special care in the reduction process would be needed in order
to cross-identify the MBP low resolution measurements of superimposed
stars with higher resolution AF and BBP measurements.}
In the AF and the BBP all regions but the very obscured dust lane
will be observable.

\begin{figure*}
\centering
\input{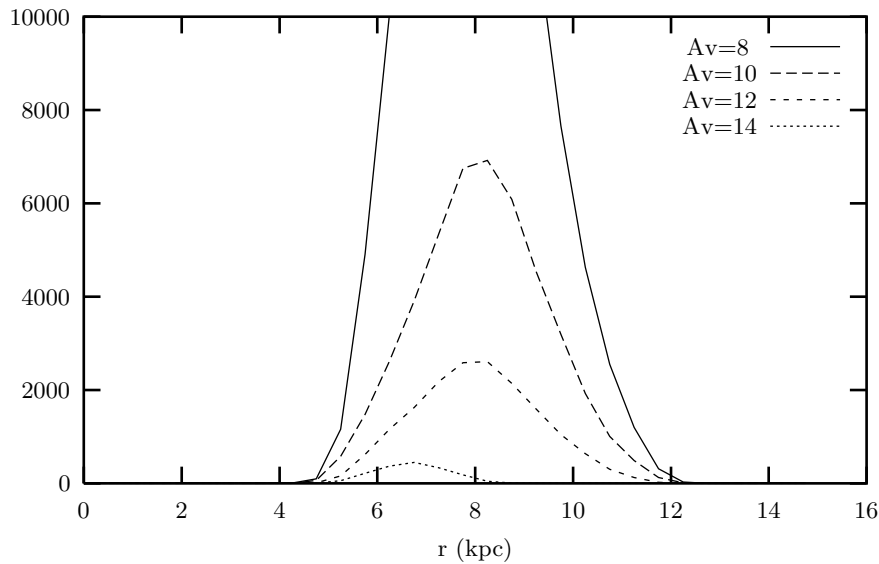}
   \caption{Estimation of the number of detected bulge stars (G$<$20)
     in the MBP as a function of distance along the line of sight in a field
at l=2$^\circ$, b=-0.5$^\circ$ with different values of visual extinction.}
   \label{fig8}
\end{figure*}

\section{Discussion and conclusions}

The method described here allows one to estimate the observability of
any critical population of the Galaxy from a variety of instruments.
The result on the bulge visibility strongly depends on the assumed extinction.
Hence the computation may be redone with more accurate maps when available. 
The bulge model used may be inaccurate in some cases (particularly
very near the Galactic center). However it has been fitted to a wide set of 
data in the near-infrared. Thus the reliability of the result is
given by the accuracy of the extinction map rather than by the
Galactic model used.
The application of the method to Gaia led to the following conclusions.

We have shown that there will
be opportunities for Gaia to access reliable measurements in the
Galactic bulge. The BBP photometer and the AF astrometer are the most 
favourable instruments as they
allow access to very dense regions of 3 million stars per square
degree.
The BBP will furnish accurate photometry in broad band filters and the
astrometer will measure accurate
parallaxes, giving an accuracy of 10\% in distance for bulge stars. 
Proper motions will also be measured in nearly all the bulge region. 
At low latitudes $|b|<1^\circ$ in the dense dust lane of the 
Galactic plane, most of the bulge stars
will be too faint (due to extinction) to be reached in a significant number
even at G$\leq$20. However in a few fields close to $b=0^\circ$, windows
of lower extinction, especially at $l>2^\circ$, bulge giants in the 
clump may be reached allowing a probe of the bulge in depth, hence a 
photometric and astrometric test of the kinematics of the outer bulge.

The MBP has a lower spatial resolution. Thus, the crowding is reached
at lower star densities and measurements will be limited to a small
range of latitudes at  $-1^\circ<b<1^\circ$, assuming the SGS map.
These estimates strongly depend on the extinction, which is not
known with sufficient accuracy. 
However, the number of accessible windows are well spread in longitude
and also in latitude (thanks to the patchiness of the extinction) and
bulge stars will also be spread in depth well inside the bulge.
Figure~\ref{fig8} shows the distribution in distance along the line of sight of
reachable bulge stars in the MBP at the extinction of A$_V$=8, 10, 12 and 
14 located
at latitude b=-0.5$^\circ$ and longitude l=2$^\circ$. The density varies 
slowly with
longitude such that these numbers should be valid  for most of the
bulge within an order of magnitude. Most of the fields at these latitudes 
have visual extinctions in the
range 8-14, with most probable values in the range 12-13.
We see in these simulations that a large number of bulge stars will be
measured in the MBP at b=-0.5$^\circ$. These bulge stars have a distance
distribution allowing one to measure the radial gradients in depth in
the bulge.

{The RVS will suffer strongly from crowding at latitudes smaller than 1$^\circ$ assuming that the} crowding
limit is really 20,000 stars per square degree. However, if 
SFD dust estimates are reliable enough at these latitudes, there may be
a significant number of small windows at various longitudes and
mostly at $2^\circ<|b|<3^\circ$ at which radial velocities of bulge giants 
will be obtained without crowding. {Even in crowded regions, spectra of bulge
giants will be obtained at G $<15$ in most bulge regions. However less extinguished regions, like the Baade's window, will suffer from too heavy crowding. 
Eventually the range of longitudes and latitudes where bulge giants will be 
measurable} will
permit us to study possible variations of radial velocities and abundances
with longitudes on each part of the dust lane of the Galactic plane.
It should be noted that the crowding limit of the RVS is not yet
accurately known. If the value used here is changed for a higher
density then a larger part of the bulge will be easily measured with the
Gaia spectrograph, in particular regions at higher
latitudes. Simulations with a crowding limit of 40,000 stars per square degree
and the SFS extinction map
show that about half of the region between 2$^\circ$ and 8$^\circ$ 
in latitude will not be crowded at G $<17$.

Putting together observations of the astrometric field, the BBP, the MBP
and the RVS, Gaia will produce a detailed survey of clump giants in the
outer bulge.
Close to the dust lane in the inner disc, a significant
number of small extinction windows will give access to a combination of
photometry, astrometry, radial velocities and spectroscopic measurements for
a large number of clump giants. Detailed analysis of these data sets
will allow us to put strong constraints on the bulge formation scenario. 
The only very obscured place unreachable with Gaia will be the region at
$|b|<0.5^\circ$ and $-3^\circ<l<2^\circ$.

\begin{acknowledgements}
We thank Gaia
  co-investigators of different instrument working groups, especially Carme Jordi and David Katz, for fruitful
  discussions during the preparation of this study. MS is supported by the 
APART programme of the Austrian Academy of Science.
\end{acknowledgements}

\end{document}